\documentstyle[12pt,a4]{article}

\def \ve#1{\mbox{\boldmath $#1$}}

\begin{document}
\thispagestyle{empty}
\vglue 0.1truecm

\begin{center}
{\Large \bf Poisson versus GOE statistics} \\
{\Large \bf in integrable and non-integrable quantum
Hamiltonians}\footnote{Submitted to Europhysics Letters} \\
\vspace*{1.2cm}
Didier POILBLANC, Timothy ZIMAN\footnote{On leave from: Department of
Physics \& Astronomy, University of Delaware, Newark DE 19716}
and Jean BELLISSARD\\
\vspace*{0.65cm}
Laboratoire de Physique Quantique\footnote{Unit\'e de recherche
associ\'e au CNRS (URA 505)} \\
Universit\'{e} Paul Sabatier \\
F-31062 Toulouse, France \\
\vspace*{0.65cm}
and \\
\vspace*{0.4cm}
Fr\'ed\'eric MILA \\
\vspace*{0.65cm}
Institut de Physique \\
1, Rue A. L. Breguet \\
CH-2000 Neuchatel, Suisse \\
\vspace*{0.65cm}
and \\
\vspace*{0.4cm}
Gilles MONTAMBAUX \\
\vspace*{0.65cm}
Laboratoire de Physique des
Solides\footnote{Unit\'e de recherche associ\'e au CNRS (URA 2)} \\
Universit\'{e} Paris--Sud \\ F-91405 Orsay, France \\
\vspace*{1.4cm}\end{center}
\newpage
{\bf Abstract:}
\par\noindent
We calculate the level statistics by finding the
eigenvalue spectrum for a variety of one-dimensional
many-body models, namely the Heisenberg chain, the t-J model
and the Hubbard model. In each case the generic behaviour is
GOE, however at points corresponding to models known to be exactly integrable
Poisson statistics are found, in agreement with an argument we outline.

\newpage
There is currently great interest in developing useful concepts
for many-body systems that go beyond perturbative methods,
particularly for low-dimensional systems where
perturbation theory has many divergences. One such concept
which has been around for some time is that of level
statistics \cite{w,d,dm,mehtabook}. The spectrum
of a many-body quantum system, originally a nucleus, in a large but finite
dimensional size Hilbert space was described in terms of a statistical
ensemble of random matrices with matrix elements of long range.
In such a statistical theory the distribution of
energy levels depends only on the symmetry of the Hamiltonian.
Later it was suggested \cite{ge,as} that the question was of interest
for the
study of small dielectric particles.
There has recently been a renewal of
attention to the question of level statistics
with the realization that they
can provide a link to semi-classical
pictures of many-body systems. The transition to chaos in the
classical system is associated with the development
of level repulsion, or, more precisely, of level statistics
characteristic of the Gaussian Orthogonal
Ensemble (GOE)\cite{bo}.
\par
In the context of single-particle Hamiltonians in a non-translationally
invariant medium the question of level statistics is relatively
straightforward : there is level repulsion provided the
eigenvectors of the infinite system are extended. If they are not,
there may be apparent level repulsion over scales small compared to the
localisation length,
but at much longer length scales states become
statistically independent\cite{loc}. For incommensurable potentials
there remain
correlations from the
potential\cite{levharp}.
\par
For many-body sytems the approach of using  level statistics to
connect the quantum behaviour to a
semi-classical description
is newer.
In an innovative paper, Montambaux et al. \cite{pioneer}
have shown that for a 2-d t--J model the level statistics
calculated for a finite lattice
give convincingly the same behaviour as the GOE.
Here we present a more systematic approach
by restricting ourselves to one spatial dimension but looking at a
variety of models starting with the Heisenberg chain, then the t--J model
and finally the Hubbard model. For each of these models there is a subset of
the parameter space for which the lattice model is integrable by the
Bethe Ansatz even for a finite chain. As it is known that integrability
correspo
the existence of an infinite number of conservation laws we may
expect \cite{pioneer},
and it turns out to be true, that the spectrum is quite different for these
cases.
Thus we may say that for the many-body system the integrable point
exhibits level statistics that correspond
to localized states in the language of a single-particle
picture, while the general points are extended. Of course the ``localisation"
here is extreme: because of the extra conservation laws corresponding to
integrability there is a
further decoupling into smaller sub-spaces.
\par
We now define the Hamiltonians considered which correspond to
increasing numbers of degrees of freedom and increasing
fermionic nature. For simplicity of
interpretion  in each case we consider models
with overall spin rotational symmetry and periodic boundary
conditions. First the Heisenberg model
\begin{equation}
H = J_1 \sum_{\bf i} \ve{S}_{\bf i} \cdot \ve{S}_{\bf i+\bf 1}
  + J_2 \sum_{\bf i} \ve{S}_{\bf i} \cdot \ve{S}_{\bf i+\bf 2}
\end{equation}
Here $\ve{S}_{\bf i}$ are spin--1/2 operators at lattice sites ${\bf i}$
of a periodic chain, ${\bf 1}$ is the unit vector and
${\bf 2}=2\times {\bf 1}$.
For simplicity we shall put $J_1 = 1$ in the following.
Increasing $J_2$ there is a transitions at  $J_2\approx 0.25$
to a spontaneously
dimerized phase. At $J_2 = 0.5$ the ground state can be written
exactly\cite{HaldaneShastry}
but the model is not believed to be completely integrable.
Now the t--J model is
\begin{eqnarray}
{\cal H} &=& J \sum_{{\bf i}}  \ve{S}_{\bf i} \cdot \ve{S}_{\bf i +\bf 1}
- t \sum_{{\bf i},{\bf\sigma}}
P_G(c^{\dagger}_{{\bf i},{\sigma}} c_{{{\bf i}+{\bf
 1}},{\sigma}} +
c^\dagger_{{{\bf i}+{\bf 1}},{\sigma}}c_{{\bf i},{\sigma}})P_G
\label {tj}
\end{eqnarray}
\noindent
where $c_{{\bf i},\sigma}$ are $hole$ operators with spin $\sigma$
and the  matrix elements are restricted
to nearest neighbor motion  on a periodic chain.
$P_G$ is the Gutwiller projector that excludes doubly
occupied sites.
The t--J model as written involves only nearest neighbour
interaction, yet it is in general believed to be non-integrable. However
for the special point J=2 there is an additional supersymmetry
between the (bosonic) spin variable and the fermions. This point of higher
symmetry, as pointed out by Sutherland, is completely integrable via
the Bethe Ansatz.\cite{supersym}

Finally we  define the Hubbard model
\begin{equation}
{\cal H} = - t \sum_{{\bf i},{ \sigma}} ( c^{\dagger}_{{\bf i},{
\sigma}}
 c_{{{\bf i}+{\bf 1}},{ \sigma}}
 + c^{\dagger}_{{{\bf i}+{\bf 1}},{ \sigma}} )
+ U \sum_{{\bf i},{ \sigma}}
 n_{{\bf i},{ \sigma}} n_{{\bf i},-{ \sigma}}
+ V \sum_{{\bf i},{ \sigma},{ \sigma^\prime}} n_{{\bf i},{
\sigma}}
n_{{\bf i}+{\bf 1},{\sigma
^\prime}}
\label{Hub}
\end{equation}
The Hubbard model is integrable for the case of on-site interaction only
($V = 0$).
\par
We find the eigenvalues in each symmetry subspace by
means of the Lanczos algorithm. In the past the primary
application of this technique to low-dimensional many
body problems  was
to study extreme parts of the spectrum since this determines the low-lying
excitations, and for scaling
studies these correspond to the longest length scales. Furthermore
the number of iterations needed is in general much smaller than the total
dimension of the Hilbert space. For the study of the complete
spectrum it is necessary to have iterations in number several times
the dimension of Hilbert space. This makes feasible calculations
restricted to systems of several thousand rather than millions in the
case of partial diagonalization with modest computational effort.
\par
As mentioned previously, we take the case of overall
rotational symmetry of the spin. This makes the
interpretation clearer but it is slightly trickier
numerically to
include total spin symmetry rather than symmetry around a single
axis.
In practice we use two different methods, both of which function
satisfactorily. One is to add to the Hamiltonian a term
$\lambda \ve{S}\cdot \ve{S}$ where $\ve{S}$ is the total spin operator.
If $\lambda$ is sufficently large each subspace of total spin separates in
energ
Discrete symmetries, for example translational and
reflection symmetries, are used
to ensure complete separation of the subspaces. For the case of the
Heisenberg chain of length $L=20$ this permits reduction of the Hamiltonian in
the most symmetric subspace (${\bf k} = 0$ , even under reflection and
spin reversal) to dimension 2518
with the subspaces
$S=0,2,4$ having 490, 1280 and 626 levels respectively (the odd spin
sectors are also odd under spin reversal). Note that since
the results are basically
independent of the choice of the quantum numbers (eg total momentum or
total spin) we shall restrict ourselves, for clarity, to subspaces with
the largest number of states.
The other method is to diagonalize in a subspace, say eg $S^z = 1/2$ (for
an odd number of particles) and then
a second time in $S^z = 3/2$. Provided the eigenvalues are complete in
each subspace, the lowest spin states are those that occur in the first space
but not the second. In the case of the t--J chain with $L=16$ sites
and  one hole
this method enabled to select the 1430 $S=1/2$ states among the
6435 states of given total momentum $\bf k$. For convenience, we choose
$\bf k$ at non-symmetric points in the Brillouin zone so that
the reflection symmetry is ineffective. However, the results
are independent of the choice of the total momentum.
Note that exactly at the integrable point $J=2$ a small proportion of
exact degeneracies appear (1215 distinct $S=1/2$ levels instead of 1430).
\par Once we have a complete eigenvalue spectrum in a given subspace
we then calculate the level statistics as previously described by
Montambaux et al\cite{pioneer}. The density of states is included
by fitting a  smooth function to a local average.
\par
We return to the argument for Poisson statistics in the case of the
integrable model. The argument here is as follows: if a Bethe Ansatz
holds,
each solution is characterised by a set of quasi-momenta $k_i$, i=1,2,..n
where, eg for an antiferromagnetic subspace n is the number of
spin-flips
relative to the ferromagnetically aligned state.
The  $k_i$, which reduce to real momenta for the non-interacting
case, are governed by a non-linear set of equations.
\par
{}From this set of equations it usually follows that the possible
quasi-momenta $k_i$ repel one another, namely they lie on
a quasi-lattice.
The energy is a sum of energies $\cos k_i$ of the quasi-momenta and
the level statistics reduces to the statistics of the lattice positions
relative to the Fermi n-sphere. Using the same argument as Berry and Tabor\cite
{BerryTabor} leads to Poisson statistics.
For comparison to the complexities of the non-interacting case see ref
\cite{Ble
\par
{}From the figures it is clear that there
is a simple conclusion in accord with the speculations by Montambaux et al.;
namely the distribution is
Poisson for the integrable cases, GOE for the generic cases\cite{clement}.
As the level distribution is observed to be independent of
the choice of subspace of total
spin $S$ we take the largest subspace calculated.
We remark that if we are interested in empirical methods of
establishing
integrability by looking at level statistics  the stronger observation is
observ
GOE. If instead
of the infinite number of subspaces of the integrable point there were a finite
number of new conservation laws at a special point (for example at the
supersymmetric point in higher dimension) we would find a law
intermediate between the GOE and Poisson. In practice however, this would
appear
more Poisson than GOE. In particular a finite superposition of
spectra, each with GOE statistics, has a non-zero probability density at zero
repulsion
$ p(0) \neq 0 $ \cite{mehtabook} and rapidly approaches Poisson.
Had we not known the integrability
of the t--J model we could probably have concluded little more than the
existenc
of degenerate multiplets.
\par
In conclusion random matrix theory, remarkably enough, works also for the
N-body
problem in quantum statistical mechanics. Even though classically the
symmetries
are identical in
the integrable or non-integrable Hamiltonians, complete integrability in the
quantum case leads to Poisson statistics
whereas the Wigner distribution for level repulsion occurs otherwise.
While Poisson statistics may not be a characteristic of integrability, a Wigner
distribution is probably a good test for absence of integrability.

\par The computer simulations were done on the CRAY--2 of
Centre de Calcul Vectoriel pour la Recherche (CCVR), Palaiseau, France.
Support from CCVR is greatly appreciated. We would like to thank
T. Hsu  and Angles D'Auriac for sending us an unpublished account
of calculations in general agreement with the numerical results (but with
different  boundary conditions) presented here. The computer programs
for the Heisenberg model were adapted from those developed by T.Z. and H.J.
Schulz, to whom we are most grateful.
We should like to thank C. Sire and M. Caffarel for useful discussions.

\end{document}